\documentclass[sn-mathphys,iicol]{sn-jnl}

\jyear{2022}%

\theoremstyle{thmstyleone}%
\newtheorem{theorem}{Theorem}
\newtheorem{conjecture}{Conjecture}

\newtheorem{proposition}[theorem]{Proposition}%

\theoremstyle{thmstyletwo}%

\theoremstyle{thmstylethree}%
\newtheorem{definition}{Definition}%
\newtheorem{lemma}[definition]{Lemma}

\newcommand\orcidicon[1]{orcid:{https://orcid.org/#1}{}}

\begin{document}

\title[A computing machinery using a continuous memory tape]{A computing machinery using a continuous memory tape}


\author*[1]{\fnm{Yigit} \sur{Oktar} }

\affil[1]{\orgdiv{Department of Computer Engineering}, \orgname{Izmir University of Economics}, \orgaddress{\city{Izmir}, \postcode{35330}, \country{Turkey}},
\orcidicon{0000-0002-8736-8013}}


\abstract{By considering a discrete tape where each cell corresponds to an integer, thus to a possible sum, a pseudo-polynomial solution can be given to subset sum problem, which is an NP-complete problem and a cornerstone application for this study, using shifts and element-wise summations. This machinery can be extended symbolically to continuous case by thinking each possible sum as a single frequency impulse on the frequency band. Multiplication with a cosine in this case corresponds to the shifting operation as modulation in communication systems. Preliminary experimentation suggests that signal generation thus solution space calculation can be done in polynomial time. However, reading the value at a specific frequency (sum value) is problematic, namely cannot be simulated in polynomial time currently. Dedicated hardware implementation might be a solution, where both circuit-based and wireless versions might be tried out. A polynomial representation is also given that is claimed to be analogous to a tape of a Turing machine. Both rational and real number versions of the subset sum problem are also discussed, where the rational version of the problem is mapped to 0-1 range with specific patterns of True values. Although this machinery may not be totally equivalent to a non-deterministic Turing machine, it may be helpful for non-deterministic universal Turing machine actualization. It may pave way to both theoretical and practical considerations that can help computing machinery, information processing, and pattern recognition domains in various ways.}

\keywords{Turing machines, non-determinism, frequency domain, Fourier transform, subset sum problem}



\maketitle

\section{Introduction}

Theory of computation involves investigation of time and space required by particular problem classes. In fact, if a problem is solvable in polynomial-time, it is deemed feasible. However, non-deterministic polynomial-time (NP) problems also exist, where the solution (if given) can be verified in polynomial time but computing the solution traditionally requires exponential time. NP-complete problems are the most difficult in NP, and all NP problems can be reduced to them in P time. This means that if one can solve an NP-complete problem in P time, it can be used to solve all of NP problems in P time~\cite{currin2017computing}.

This paper proposes a novel computing machinery within the frequency domain for solving the subset sum problem, which is proven to be of class NP-complete. NP-complete problems are assumed to be computationally hard to solve on conventional electronic computers, namely deterministic universal Turing machines (or computers with von Neumann architecture in practice). Subset sum problem is a typical problem from this class. It is a decision problem where given a list of positive integers one needs to check whether any subset acquires a target sum value. The most naive solution would check the sum of all possible subsets, thus be of exponential time complexity. There is no known polynomial time solution for this problem. However, there are pseudo-polynomial solutions. In this paper, we first provide a pseudo-polynomial solution and then through further continuous generalization of that solution, we sketch a frequency domain approach to solve subset sum problem.

\begin{definition}[SUBSET SUM]
Given a list of $n$ natural numbers $A = \{a_{1},....,a_{n}\}$, the problem is to decide whether any subset of these integers sums precisely to a given target value $T$.
\end{definition}

\begin{definition}[SUBSET SUM: INTEGER VARIANT (IV)]
The inputs in $A$ are integers.
\end{definition}

\begin{definition}[SUBSET SUM: ZERO-SUM VARIANT (ZSV)]
The inputs in $A$ are integers, and $T=0$.
\end{definition}

It is proven that the SUBSET SUM problem is NP-Complete~\cite{kleinberg2005algorithm}. IV and ZSV are variants of this definition and are also known to be equivalent to the original definition in terms of time complexity. 

There have been attempts to build non-deterministic machinery to solve NP-complete problems, through photons~\cite{xu2020scalable}, or molecular consideration to name a few~\cite{henkel2007dna}. In this study, the frequency band is regarded as a continuous tape and using modulation, parallel writes can be performed on this tape corresponding to virtually infinite number of heads. Therefore, a non-deterministic computing machinery can be developed within the frequency domain. Such an approach to computation might pave way to deeper theoretical insight in the field of computer science. For example, if simulation of this hypothesized machine can be performed using a regular computer efficiently (in polynomial time), it may provide insight for one of the millennium problems namely, whether P is equal to NP and it will have drastic consequences. 

Even if efficient software simulation is not possible, this machinery will count as a type of non-deterministic computer and a dedicated hardware implementation would still solve the subset sum problem faster than a deterministic one.

\section{Previous work}

In simple terms, the Turing machine model uses an infinite discrete tape as its unlimited memory. In fact, Alonzo Church and Alan Turing equivalently defined what an algorithm means, Church with $\lambda$-calculus, and Turing with his machines~\cite{turing_1937}. Their definition of an algorithm helped to resolve Hilbert's tenth problem, namely testing of a polynomial to have integral roots, and gradually the limits of computability. Also, the notion of Universal Turing Machine proposed by Alan Turing helped the development of stored-program computers, namely conventional computers. The practical realization of a UTM is performed through von Neumann architecture~\cite{von1993first}. It can be viewed as a device that has a central processing unit (CPU), that is physically separate from the memory. This CPU has a control unit to direct the sequential operation and an arithmetic/logic unit (ALU) that contains all the arithmetic functions and logic gates that are needed. However, this type of approach requires a large amount of data to be transferred back and forth between the CPU and the physically separate memory, leading to the 
concept of von Neumann bottleneck, both in time and energy~\cite{traversa2015universal}. 

With respect to the subset sum problem, previous work can be categorized in three branches. First, efficient solutions using deterministic Turing machines can be given. For example, there is a pseudo-polynomial approach based on dynamic programming with a run-time of $O(sN)$ where $s$ is the desired sum and $N$ is the number of elements. In 2015, Koiliaris and Xu found a deterministic ${\displaystyle {\tilde {O}}(s{\sqrt {N}})}$ algorithm for the subset sum problem where ${\displaystyle s}$ is the sum we need to find.~\cite{koiliaris2017faster} In 2017, Bringmann found a randomized ${\displaystyle {\tilde {O}}(s)}$ time algorithm~\cite{bringmann2017near}. However, in a typical hard case integers in the set might have values as high as $2^N$ thus $s$ might as well be exponential in input size. 

On the other side, as the second branch, dedicated hardware implementations of non-deterministic Turing machines are considered. Universal memcomputing machines, a class of brain-inspired general purpose computing machines power-wise equivalent to non-deterministic TMs have been proposed ~\cite{traversa2015universal}, showing similarities to our approach. Other approaches include molecule-based computation using large quantities of DNAs ~\cite{henkel2007dna}. Photonic investigation is also promising for non-deterministic computation~\cite{xu2020scalable}.

As the last branch, there is also the quantum computation model, though totally not equivalent to non-determinism. Current studies in this branch can only claim faster exponential-time algorithms~\cite{bernstein2013quantum}, or polynomial time only when certain conditions are met~\cite{daskin2017quantum}. Through the ability of a post-selected quantum computing, which is an imaginary computing model, one can obtain the result of a Grover search problem in O(1)~\cite{farhi2016quantum,aaronson2005quantum}. This would lead a quantum computer to solve NP-complete problems in polynomial time. Although this model is imaginary, it still provides insights for quantum and classical computers on the solution of NP-complete problems~\cite{daskin2017quantum}. Namely, a quantum computer can generate all the solution space and mark the correct answer in polynomial time. However, this information, the marked item, can only be obtained by an observer with an exponential overhead~\cite{daskin2017quantum}. 

In this paper, we present a model of computation that can also generate the solution space in polynomial time. However, reading the solution is still problematic, which lead us to consider dedicated hardware implementation to perform also the reading efficiently.
\section{Proposed Methodology}

The origination point of the proposed method is the subset sum problem. The idea is to keep track of a tape that has a cell for each positive integer. The value within $X^{th}$ cell then means that the target sum of $X$ is attainable that many ways. Let us formulate the algorithm in detail in the upcoming subsection.   

\begin{figure}
\includegraphics[width=8cm]{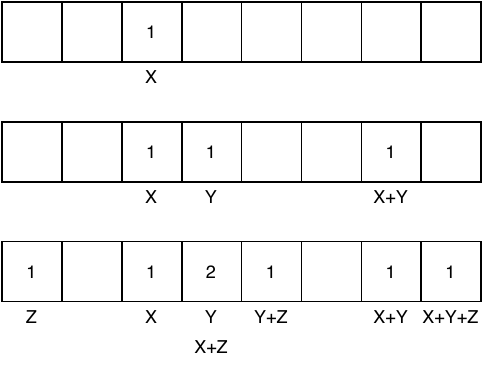}
\caption{For the case of 3 values (X=3, Y=4, Z=1), by simple shifts and element-wise summation all combinations are calculated on the tape in 3 steps}
\label{1}
\end{figure}

\subsection{Pseudo-polynomial algorithm}

Assume that we have a discrete tape of cells as in a Turing machine. Assume that our first value is $X$ (i.e. a[0] in Algorithm~\ref{algo1}). In that case, we mark the $X^{th}$ cell on the tape from left with 1. At each iteration, we copy the tape and shift it with an amount equal to new value, also increase value in this new corresponding cell position in this new tape by 1. When the old tape is summed with this new tape in an element-wise manner, the resulting nonzero values of tape designate the possible sums and their number of ways. Namely, if $tape[i]=p$, a sum of $i$ is attainable in $p$ many ways. Pseudo-code of this algorithm is given in Algorithm~\ref{algo1}. A depiction of steps for the algorithm is shown in Fig.~\ref{1}.
The running time for this algorithm is $O(SN)$ where $S$ is the sum of all array values and $N$ is the number of array elements.

\begin{algorithm}
\label{algo1}
 S = sum of all array values; \\
 int[] tape = new int[S+1]; \\
 tape[a[0]] = 1; \\
 for(int i=1; i$<$ a.length; i++)\{ \\

~~int[] tape2 = new int[2*S+1];\\ 

~~for(int j=0; j$<=$S; j++)\{\\
~~~~tape2[j+a[i]]=tape[j];\\
~~\}

~~tape2[a[i]]+=1;\\
  	
~~for(int j=0; j$<=$S; j++)\{\\ 
~~~~tape[j] = tape[j] + tape2[j];\\
~~~~\}
\}
	    		
\caption{A pseudo-polynomial time solution to the subset sum problem}
\end{algorithm}

Note that, in Algorithm~\ref{algo1} all multiplicities of sums are tracked. For the decision version of the problem, instead of keeping multiplicities one can keep only a binary tape (true if the sum is possible, and false otherwise), instead of an integer tape. 

\subsection{Extension to continuous case}
\label{continuous_case}
It is possible to extend this concept into continuous case. In this case we attain a continuous frequency band. Shifting operation then is sustained via modulation with a sinusoidal as in communication systems. If all calculation is symbolic, continuity is assured. 

\subsubsection{Signal generation}
\label{signal_gen}
When a value arrives, we first make it a monotone signal through forming a sinusoidal with that specific frequency. In other words, it reflects a single frequency (an impulse) within the frequency domain. As new value(i.e. X) arrives, we shift the old band by multiplication with $cos(2*\pi*X*t)+j*sin(2*\pi*X*t)$ and sum it with the previous version. Last line of Equation~\eqref{rule} indicates this process in mathematical form. Equation set~\eqref{sn} indicates $S_{N}(t)$ as a product of $N$ terms.

\begin{equation}
\begin{split}
S_{0}(t) & = 1 \\	
S_{i}(t) & = S_{i-1}(t)(1+e^{j 2\pi a[i-1] t})
\end{split}
\label{rule}
\end{equation}

\begin{equation}
\begin{split}
S_{1}(t) & = S_{0}(t)(1+e^{j 2\pi a[0] t}) \\
S_{1}(t) & = (1+e^{j 2\pi a[0] t}) \\
S_{2}(t) & = S_{1}(t) (1+e^{j 2\pi a[1] t}) \\
S_{2}(t) & = (1+e^{j 2\pi a[0] t}) (1+e^{j 2\pi a[1] t}) \\
... \\
... \\
... \\
S_{N}(t) & = \prod_{i=0}^{N-1} (1+e^{j 2\pi a[i] t})
\end{split}
\label{sn}
\end{equation}

\subsubsection{Reading the sum value using Fourier transform}
Signal generation, namely acquiring $S_{N}(t)$ can be done in polynomial time in symbolic form when it is represented as a product of $N$ terms. However, reading the desired value at a specific frequency is more problematic.  To check whether a sum of $b$ exists, we need to query whether the Fourier transform of $S_{N}(t)$ has a recording at $2\pi b$ or not. The Fourier transform of $f(t)$ with respect to variable $t$ at the point $w$ is

\begin{equation}
\mathcal{F}\{f(t)\} = F(w) = c \int_{-\infty}^{\infty} {f(t)e^{iswt}dt} \\
\end{equation}

where $c$ and $s$ are parameters of the Fourier transform. Parameter $c$ is the normalization parameter and is set to $1$. While $s$ is the oscillatory factor and is set to $-1$. If the final line of equation \eqref{sf} is non-zero then the sum $b$ exists within the signal. The whole process is simulated in Algorithm~\ref{algo2} in Matlab.

\begin{equation}
\begin{split}
\mathcal{F}\{S_{N}(t)\} & = \mathscr{S}_{N}(w)  \\	
\mathscr{S}_{N}(w) & = \int_{-\infty}^{\infty} \{ {\prod_{i=0}^{N-1} (1+e^{j 2\pi a[i] t}) \} e^{-j w t} dt} \\
\mathscr{S}_{N}(2\pi b) & = ? \\
\end{split}
\label{sf}
\end{equation}

At this point, for time complexity analysis purposes, we can go back to Equation~\eqref{sn} and realize that we can write $S_{N}(t)$ as a sum like

\begin{equation}
S_{N}(t) = 1+\sum_{i=0}^{M-1} {e^{(j 2\pi b[i] t)}},
\end{equation}

where $b[i]$ include each possible sum value that every combination of $a[i]$ provides, and $M$ the number of possible sums. Here the important thing to notice is that $N \leq M \leq 2^{N}-1$. In the worst case, every combination of $a[i]$ may result in a distinct sum value of $b[i]$ thus resulting in $M=2^{N}-1$.

Writing $S_{N}(t)$ as a sum of terms then helps us further when taking the Fourier transform. Due to linearity of Fourier transform Equation~\eqref{sf} then can be rewritten as

\begin{equation}
\begin{split}
\mathscr{S}_{N}(w) = \mathcal{F}\{1\} + \sum_{i=0}^{M-1} \mathcal{F}\{e^{j 2\pi b[i] t}\} \\
\mathscr{S}_{N}(w) = 2 \pi \delta (w) + \sum_{i=0}^{M-1} {2 \pi \delta (w-2\pi b[i])} \\
\end{split}
\label{nft}
\end{equation}

Here $\delta (w)$ denotes the impulse function. Notice that, $\mathscr{S}_{N}(w)$ is non-zero at $2 \pi b$ (desired sum value) if an impulse of $\delta (w-2\pi b)$ exists within the sum in Equation~\eqref{nft}. The important thing to notice is that, Equation~\eqref{nft} urges us to take Fourier transform of $M$ many terms and $M$ can be exponential in $N$. Therefore this suggests that by explicitly using Fourier transform we cannot solve the problem in polynomial time. It would be better if we could read the value in a more efficient manner.

\begin{algorithm}
\label{algo2}
syms  t;\\
n = 5; \\
S = 1; \\
a = randi($2^n$,1,n);\\
for i=1:numel(a)\\
S = S*(1+(cos(a(i)*2*pi*t)+1j*sin(a(i)*2*pi*t)));\\
end\\
tf = 9;\\
F = fourier(S);\\
eval(subs(F,2*pi*tf))\\
\caption{A Matlab script to symbolically calculate the corresponding frequency band and then read the value at 9, namely check whether sum 9 exists within this n sized random array.}
\end{algorithm}

\subsubsection{Reading the sum value through convolution}

Another reading mechanism in software is possible through symbolic convolution. We can use the fact that convolution in time domain corresponds to multiplication in frequency domain. Convolution of two functions $f(t)$ and $g(t)$ is given as

\begin{equation}
[f * g](t) = \int_{-\infty}^{\infty} {f(\tau)g(t-\tau) d \tau}
\end{equation}

Therefore, if we convolve the signal $S_{N}(t)$ with $e^{j 2 \pi b t}$, it would then correspond to multiplication of $\mathscr{S}_{N}(w)$ with desired impulse function $\delta (w-2\pi b)$ and if the resulting value is non-zero it means that the signal includes a sum value of $b$.

\begin{equation}
\int_{-\infty}^{\infty} (\prod_{i=0}^{N-1} (1+e^{j 2\pi a[i] \tau}))(e^{j 2 \pi b (t-\tau)})d \tau
\label{conv}
\end{equation}

However, performing this symbolic convolution in time domain may not be polynomial still as Equations~\eqref{sf} and~\eqref{conv} are very similar in nature. Although we cannot read the value efficiently in software, it is claimed that when the signal $S_{N}(t)$ is generated physically, then an efficient physical reading can be performed with a dedicated hardware solution utilizing the concepts of communication systems. When done physically, there is no need to take a Fourier transform, but one can easily tune in and read whether there is content at a specific frequency.

\subsection{Experimentation}

An approximate average case run-time analysis of the algorithm in Section~\ref{signal_gen} is now given. Maximum possible integer is set to $2*10^{9}$ for random number generation of the array values. A set of 5 tasks are performed for each array of sizes 1000 to 10000 with varying length of 1000, each array value artificially generated with the random number generator mentioned. Then mean run-time of each array size is calculated from 5 of these for each array length. This gives an approximate average run-time analysis. Namely, we can depict the average run-time with varying sizes of array length as in table~\ref{avg_run_time}. Using the \textit{polyfit} function of Matlab on these values shows an at most quadratic relationship between the array size and the average case running-time. However, a worst-case time complexity analysis must be preferred for more to comment on this issue. Experimental worst-case time analysis is not feasible as one needs to find the worst case example for each case through a humble combinatorial search (which makes the naive search unfeasible). Theoretical worst-case time analysis of signal generation is left as a future work on this occasion.

\begin{table}[]
\caption{Average case run-time analysis of the signal generation process.}
\label{avg_run_time}
\begin{tabular}{ll}
\hline
\textbf{Array length}      & \textbf{Time to run (seconds)}        \\ \hline
1000         & 8.08        \\ \hline
2000         & 26.38       \\ \hline
3000         & 57.37       \\ \hline
4000         & 99.54       \\ \hline
5000         & 154.09      \\ \hline
6000         & 231.60      \\ \hline
7000         & 307.73      \\ \hline
8000         & 401.72      \\ \hline
9000         & 512.43      \\ \hline
10000        & 650.58        \\ \hline
\end{tabular}
\end{table}

\section{Notes on dedicated hardware implementation}

An efficient software simulation of the proposed method has not yet been achieved, namely signal generation (or calculating the solution space) can be done in polynomial time; however, reading the values is not easy on software as it requires to symbolically calculate the Fourier transform or perform a symbolic convolution. Therefore, a dedicated hardware implementation might be useful from both theoretical and practical perspectives. In this regard, there are two possible approaches for dedicated hardware implementation. First, circuit based implementation and the second is wireless implementation. Wireless implementation might be straightforward as communication system principles apply. For circuit-based implementation, an equivalent formulation of the problem from a different perspective is discussed.

\subsection{Circuit-based implementation}

First of all, subset sum problem can be converted to a problem where the array elements can be both positive and negative integers. Then, the question can be whether there is a subset that sums up to zero or not. In continuous frequency band formulation, this corresponds to whether there is extra zero frequency component at the end of calculation (besides the component that comes from $S_{0}(t)=1$ at initialization).

\begin{equation}
\begin{split}
\int_{t=0}^{t=1} S_{N}(t) dt & = \int_{t=0}^{t=1}  \prod_{i=0}^{N-1} (1+e^{j 2\pi a[i] t}) dt \\
& = \int_{t=0}^{t=1} (1+\sum_{i=0}^{M-1} {e^{(j 2\pi b[i] t)}}) \\
& = \int_{t=0}^{t=1}{ 1 dt}+ \sum_{i=0}^{M-1} {\int_{t=0}^{t=1} {e^{(j 2\pi b[i] t)} dt}}
\end{split}
\label{integ}
\end{equation}

Existence of extra zero frequency component is equivalent to the existence of extra direct current within the signal. Assuming that we are using integer frequencies, this then corresponds to checking whether the integral of our cumulative signal function $S_{N}(t)$, in the interval from 0 to 1, is more than 1 or not. If the integral is 1 there is no extra direct current thus there is no sum to 0, if the integral is above 1, then there has to be a direct current component and equivalently a subset that sums to zero. The integral of the total signal will be above 1 if and only if there is an extra zero frequency component. Therefore, the subset sum problem boils down to the detection of an integral above 1 in time domain in this case from $t=0$ to $t=1$ as formulated in Equation~\eqref{integ} and Equation~\eqref{zero}.

\begin{equation}
\begin{split}
{\int_{t=0}^{t=1} {e^{(j 2\pi b t)} dt}} & = -\frac{j(-1+e^{j b 2\pi})}{2 \pi b} \\
 & = 0, b = n, n \neq 0, n \in Z \\
 & = 1, b = 0 \\
\end{split}
\label{zero}
\end{equation}

\subsection{Wireless implementation}
Radio is the technology of signaling and communicating using radio waves~\cite{ellingson2016radio}. These waves are electromagnetic waves of frequency between 30 hertz and 300 GHz or $3\times10^{11}$ hertz. In our formulation, we used integer frequencies. However, scaling can be performed to map the solution space into this 30 Hz - 300 Ghz range, if radio waves are to be used. Microwaves can also be used after a correct scaling. The initial task is to generate $S_{N}(t)$ and broadcast it (though only locally). The most important and impractical fact is that the space should be eliminated of previously generated electromagnetic waves in the air as they all are unnecessary signals in the eyes of our application. This may be possible through the usage of a additional hardware. The other possible solution is to use a spectrum which is not being used by other applications. Note that, in all cases electromagnetic waves are to be processed only locally.  

After generating and publishing the signal into a proper local medium, the only task that remains is to check the occurrence of the desired frequency. Depending on the scaling factor $\gamma$, this can be performed by bandpass filtering. Assuming that the sum value we are interested in is $b$ we can form a bandpass filter within the range $[b/\gamma-\epsilon,b/\gamma+\epsilon]$, where $\epsilon$ depends on the scaling factor $\gamma$, namely $\epsilon < 1/\gamma$. After filtering, if we acquire a non-zero signal, or more practically a signal higher than a certain predetermined threshold, it means that the sum $b$ occurs. It is claimed that this procedure will be far more effective than explicitly computing the Fourier transform of $S_{N}(t)$ symbolically and querying the desired frequency.

\section{More theoretical viewpoint}\label{sec1}

\subsection{Polynomial representation of a tape of a Turing Machine}
\label{poly}

Formal power series are often used to describe infinite sequences through the usage of generating functions. Therefore, it is possible to simulate an infinite tape (of a Turing machine) through the usage of generating functions where each element of the underlying sequence corresponds to a cell of the original tape, namely $a_{n}$ representing the value of $n^{th}$ cell. One can assign each integer modulo $k$ to a symbol occurring in the tape alphabet of size $k$ of the underlying Turing machine. This way, a direct correspondence can be built between an infinite polynomial representation and an infinite tape.

\begin{definition}

The ordinary generating function of a sequence $a_{n}$ is $G(a_{n},x) = \sum_{n=0}^{\infty}{a_{n}x^{n}}$

\end{definition}

\begin{definition}

An infinite polynomial tape of alphabet of size $k$ is the ordinary generating function of any possible sequence $a_{n}$ where $a_{n} \in \Z{k}$, denoting the group of integers modulo $k$. In that case, zero terms of the generating function correspond to blank cells of the tape.

\end{definition}

Having defined a polynomial representation of a tape, it is better to give necessary functionality so that the tape is useful. Namely, it is better to explain how reading from and writing onto this tape is to be performed.

\begin{lemma}
\label{reading}
Let $F(x)$ designate the current status of the polynomial tape. Then, the value of $n^{th}$ cell can be read by performing $\frac{d^{n}{F(x)}}{dx^{n}}\Big\rvert_{x=0}\frac{1}{n!}$, namely taking the $n^{th}$ derivative at $x=0$ and dividing by $n$ factorial. 
\end{lemma}

\begin{proof}[Proof of Lemma~\ref{reading}:]
$F(x) = P(x) + a_{n}x^{n}+  Q(x)$, let us divide $F(x)$ into 3 parts, namely to $P(x)$, $a_{n}x^{n}$, and $Q(x)$, where $P(x)$ includes the monomials with power less than $n$ and $Q(x)$ with power more than $n$. Then $\frac{d^{n}{F(x)}}{dx^{n}}\Big\rvert_{x=0} = \frac{d^{n}{P(x)}}{dx^{n}}\Big\rvert_{x=0} + \frac{d^{n}a_{n}x^{n}}{dx^{n}}\Big\rvert_{x=0} + \frac{d^{n}{Q(x)}}{dx^{n}}\Big\rvert_{x=0}$ The first term, $P(x)$ vanishes as it includes the monomials with power less than $n$. The second term evaluates to $n!a_{n}$ at $x=0$ and third term $Q(x)$ also vanishes as at $x=0$ because the derivative includes extra $x$ powers. Therefore, through $\frac{n!a_{n}}{n!} = a_{n}$, we are able to recover $n^{th}$ value of $a$.
\end{proof}

\begin{lemma}
\label{shifting}
Let $F(x)$ designate the current status of the polynomial tape. Then, the tape can be shifted $t$ many cells to the right by performing $F_{new}
(x) = x^{t}F(x)$. Note that, shifting to the left can analogously be performed by $x^{-t}F(x)$, but this may introduce negative exponents at the start of the tape invalidating generating function logic. Note that, depending on previous right shift amount, it is possible to shift left without invalidating the logic. This lemma can be used to simulate the head of a Turing machine.
\end{lemma}

\begin{proof}[Proof of Lemma~\ref{shifting}:]
Let us expand $F_{new}(x) = x^{t}F(x)$. $F_{new}(x) = x^{t}(a_{0}x^{0}+...+ a_{i}x^{i} + ... + a_{S}x^{S})= a_{0}x^{t}+...+ a_{i}x^{i+t} + ... + a_{S}x^{S+t}$. The content of old $x^{i}$ is same with new $x^{i+t}$. Note that, the new cells of $x^{0}$ to $x^{t-1}$ are initiated as blank. 
\end{proof}

\begin{lemma}
\label{writing}
Writing onto the tape can be performed by first reading the value at the $i^{th}$ cell (using Lemma 6) and suppressing it (extracting it with a negated value) and then entering the new value from the alphabet for the $i^{th}$ cell (namely adding $F(x) = F(x) + a_{i}^{new}*x^{i}$ where $a_{i}^{new}$ is the new value.
\end{lemma}

\begin{proof}[Proof of Lemma~\ref{writing}:]
Using Lemma~\ref{reading} we can extract the value at $i^{th}$ cell, namely $a_{i}$. Then, we can define an intermediate state where the original value is erased from the tape with $F_{inter}(x) = F(x) - a_{i}x^{i}$. Then the final state of the polynomial tape is $F_{final}(x) = F_{inter}(x) + a_{i}^{new}x^{i}$. Together as a single step, $F_{final}(x) =   F(x) - a_{i}x^{i} + a_{i}^{new}x^{i}$. Using $P(x)$ and $Q(x)$ as in Lemma~\ref{reading}. $F_{final}(x) = F(x) -a_{i}x^{i} + + a_{i}^{new}x^{i} =  P(x) +a_{i}x^{i} + Q(x) - a_{i}x^{i} + a_{i}^{new}x^{i} = P(x) +a_{i}^{new}x^{i} + Q(x)$. In simple terms,  $a_{i}x^{i}$ in $F(x)$ is replaced with  $a_{i}^{new}x^{i}$.
\end{proof}

It is claimed that using these lemmas it is possible to simulate a Turing machine using such a polynomial representation as its tape. The settlement of the conjecture ~\ref{poly_turing} is left as an exercise for the reader. However, the efficiency of such a simulation is still in question. On the other hand, it is possible to adjust this polynomial representation and target the subset sum problem in a more specific way. 

\begin{conjecture}
\label{poly_turing}
One can simulate a deterministic Turing machine using the polynomial representation with Lemmas 6,7,8.
\end{conjecture}

\subsection{Polynomial Representations for the Subset Sum Problem}
\label{poly_subset}
The polynomial representation to be used in this paper is a generalization of the method presented in~\cite{7029665} and a simplification of the generating function method presented in~\cite{weisstein_subset_nodate}. Assuming original definition, namely the input $A$ consists only of natural numbers, we can express the possible subset sums and their multiplicities through a polynomial expression given as,

\begin{equation}
\label{expansion}
G(x) = \prod_{i=1}^{n}{(1+x^{a_{i}})} = 1+\sum_{j=1}^{m}{c_{j}x^{b_{j}}}
\end{equation}

where $m$ designates the number of possible different sums, $b_{j}$ a specific sum, and $c_{j}$ the multiplicity of the given specific sum $b_{j}$. For simplicity purposes, without loss of generality, we can assume that $\{b_{1},..,b_{m}\}$, namely the list of all attainable sums, are sorted in ascending order. To answer the subset sum problem, it is only needed to check whether the term $x^T$ appears in the expanded form, namely there exists $b_{j*}=T$ and $c_{j*}>0$ where $j*$ designates the needed index assuming that the sum exists. 

In this way, a bridge between the subset sum problem and univariate polynomial identity testing can be built. An efficient deterministic solution may not expand the product to explicitly calculate all these $m$ terms (because there may be $2^n$ distinct sums in the worst case), but should perhaps try to extract the solution, namely the coefficient of $x^T$, in a more concise way. A straightforward way to check the existence of $x^T$ is to check whether the $T^{th}$ derivative of $F(x)$ at $x=0$ is nonzero (which is also the direct consequence of Lemma 6, see also Proposition~\ref{get_t}). Worst-case time complexity of such a step is still exponential as there can be exponential many terms (or possible sums) within the expression.

\begin{proposition}
\label{get_t}
There exists a subset of sum of $T$ if and only if $\frac{d^{T}{G(x)}}{dx^{T}}\Big\rvert_{x=0} \gt 0$
\end{proposition}

\begin{proof}[Proof of Proposition~\ref{get_t}:]
Proof is similar to Lemma~\ref{reading}. We read the value and if it is non-zero than that means $a_{T}$ is non-zero. Namely, the $T^{th}$ derivative of $P(x)$ vanishes. Extra powers exist for $T^{th}$ derivative of $Q(x)$, so evaluated at $x=0$, $Q(x)$ also vanishes. The only left value is $T!a_{T}$ if and only if the term $x^{T}$ originally exists or in other words sum $T$ occurs. 
\end{proof}

At this point, the formulation here is to be further generalized through the logic of a possible continuous tape instead of a discrete one. Advantages of a continuous tape is to be investigated further.

\subsection{Continuous generalization of a memory tape}
Note that, by allowing negative exponents the polynomial representation can be generalized. Moreover, one may allow real numbers instead of integers to further generalize the problem. Namely, the underlying polynomial representation then would have real valued exponents. A frequency domain approach is already present for the continuous case and the polynomial representation is dropped and already developed frequency domain approach is adopted as in Section~\ref{continuous_case}.

\section{Further applications}
\label{app}

\subsection{Subset sum with rational or real numbers}

Subset sum with rational numbers (with the Puiseux series correspondence for the case of a polynomial tape) instead of integers  is already a well defined problem. Note that, this version of the problem is proven to be strongly NP-complete~\cite{wojtczak2018strong}. It means that there are no possible pseudo-polynomial algorithms for this version of the problem, and if there are any then it shows P=NP. Note that, the subset sum in its original form is classified as a weakly NP-complete problem, meaning that there exist pseudo-polynomial algorithms to solve it. 

One can further define the subset sum problem over the real numbers. It is unclear whether such a problem will be decidable as there exist uncomputable numbers as the subset of real numbers (i.e. Busy Beaver function). A further parallel has to be drawn with the halting problem to classify this version of the problem with real numbers. This line of generalization then can even include complex numbers, where the represented tape then becomes a 2D tape (or the polynomial representation of the tape is then multivariate).

Note that a special 2D tape, such as the one used in the proof of rational numbers being countable, can be used in a possible pseudo-polynomial algorithm. The existence of a pseudo-polynomial algorithm to rational version of the problem would prove that $P=NP$, therefore probably such an algorithm does not exist.

\subsection{Binary version of the tape is enough}

\label{binary}
Instead of keeping the multiplicities of each possible sum, a single Boolean variable can be kept for each target value denoting whether the value is possible or not.
In natural discrete case, then each input array can be mapped to an positive integer value denoting the decimal value of the final representation of the tape. This is a concise representation of all possible sum values of the input array. In other words, given the integer value one can decide whether a sum exists or not by querying the binary representation of this integer value. Using this logic one can include references to certain OEIS (Online Encyclopedia of Integer Sequences) entries. Such investigation is left as an exercise for the reader.

\subsection{Continuous binary tape}

Binary version of a continuous tape may shed light to rational and real number variants of the subset sum problem. Namely, all of the possible sums might be mapped to 0 to 1 range if the input is scaled by $S$, the maximum possible sum. Then, if the original subset only consists of natural numbers, then the scaled version will be a rational number problem. Scaled version of the tape generates patterns of true values scattered discretely among the range of 0 to 1. Such a consideration of a continuous binary tape might be related with strong NP-completeness of rational version of the problem. There still exists a possibility of a pseudo-polynomial solution for the rational number version based on this logic.

\subsection{Polynomial identity testing}

Polynomial identity testing (PIT) is the problem of efficiently determining whether two polynomials are identical. Univariate version of PIT, namely UPIT asks whether two polynomials $P(x)$ and $Q(x)$ (with degree $d$) are identical. This is the same as asking whether $P(x)-Q(x)$ is the zero polynomial. UPIT has similarities with the polynomial representation of the subset sum problem as in Section~\ref{poly_subset}. A naive solution to UPIT would be to expand the expressions $P(x)$ and $Q(x)$ and check whether they have the same monomials in their expanded forms. A more efficient solution exists if we can show that $Z(x) = P(x)-Q(x) = 0$, namely whether $Z$ is the zero polynomial. In that case, we may not need to consider expansion of expressions, but evaluation of $Z(x)$ to $0$ at $d$ many distinct points suffice~\cite{künnemann2018nondeterministic}. Applying this to our problem, optimistically, for the polynomial representation of the subset sum problem, expansion of the expression as in Eqn.~\ref{expansion} may not be necessary.  There may be a way to figure out the existence of a certain monomial (thus a solution to the subset sum problem) without expanding all of the terms. However, a similar solution would still count as a pseudo-polynomial algorithm if it includes $d$ as $d$ corresponds to the maximum possible sum $S$. There may be more efficient solutions.

\subsection{Artificial neural networks}
\begin{figure}
\label{clnn}
  \caption{A neural network architecture consisting of three input and two output nodes and that includes a continuous memory tape as its hidden layer}
  \centering
    \includegraphics[width=0.25\textwidth]{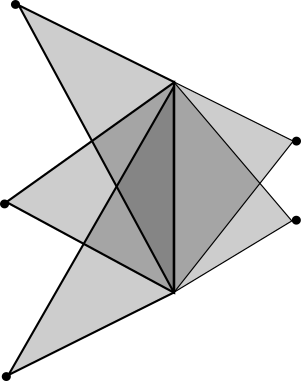}
\end{figure}

A continuous line of values can be utilized as a hidden layer of a neural network as exemplified in Fig.~\ref{clnn}. The details of such an entity are already discussed in previous works~\cite{pmlr-v2-leroux07a,Oktar2022}. Note here that, the values in the hidden layer and the values as weights can be continuous functions in this case as opposed to discretely placed values within a continuous range which we come across in the solution of subset sum problem. This opens up the possibility of intervals of values and interval arithmetic logic, though possibly not needed for the subset sum problem. Moreover, preliminary experimentation suggests that these architectures posses extensive time complexity.

\section{Discussion}

The most important discussion will be around the symbolic simulation case as in Algorithm~\ref{algo2}. The experimentation suggests that signal generation in Algorithm~\ref{algo2} can be performed in polynomial time. However, when querying a solution, we need to check whether that exact frequency exists or not. In Algorithm~\ref{algo2} this is performed by symbolic \textit{Fourier} function. However, the time complexity of this function may not be polynomial. It is claimed that if there exists a polynomial time simulation of this symbolic process, it provides a polynomial time solution to a NP-complete problem and may be the algorithm to prove P=NP.

If polynomial time simulation of this process is not possible, still there can be a dedicated hardware architecture solution. This may then count as a successful implementation of a computing machinery specifically designed to solve the subset sum problem. In this case, the dedicated machine might be a circuit based system operating on direct and alternative current, or a wireless device utilizing communication system principles.

Methodology of this paper is general enough that a subset sum problem over the rationals~\cite{wojtczak2018strong} or even reals can also be properly targeted upon further consideration. Also, the machinery can be reconsidered from the perspective of $\lambda$-calculus, a Turing complete model of computation. Namely, non-deterministic versions of $\lambda$-calculus might help drawing further parallels between\cite{10.1007/978-3-540-92687-0_8}. 

\section{Conclusion}
The central contribution of this paper is the possibility of deterministic simulation of proposed machinery in polynomial time. This problem is not resolved. However, as von Neumann architecture is the dominating implementation of deterministic universal Turing machines, there is no consensus on hardware specifications for non-deterministic universal Turing machines. A machinery within the frequency domain has been presented. Although this machinery may not be totally equivalent to a non-deterministic Turing machine, it may be helpful for non-deterministic universal Turing machine actualization. It may pave way to both theoretical and practical considerations that can help computing machinery, information processing, and pattern recognition domains in various ways.

\section*{Acknowledgments}

Author is grateful to Prof. Dr. Cem Evrendilek for fruitful discussions, and for his constructive comments that greatly improved the manuscript. Author is also grateful to his elder brother Ya\u{g}{\i}z Oktar for encouraging himself to learn and love mathematics starting from the primary school.



\bibliography{sn-bibliography}


\end{document}